\documentclass[aps,prd]{revtex4}
\usepackage{subfigure,graphics} 
\usepackage{flafter} 
\begin{document} 

\title{
Thermal Effects on Pure and Hybrid Inflation} 
\author{Lisa M. H. Hall}
\email{lisa.hall@ncl.ac.uk}
\author{Ian G. Moss}
\email{ian.moss@ncl.ac.uk}
\affiliation{School of Mathematics and Statistics, University of  
Newcastle Upon Tyne, NE1 7RU, UK}

\date{\today}


\begin{abstract}
This paper discusses models of inflation based on global supersymmetry. It is
shown that there are parameter ranges, consisent with observational
constraints, for which warm inflation occurs and supergravity effects can be
neglected. There is no need for any fine tuning of parameters. The thermal
corrections to the inflaton potential are calculated and it is shown that they
do not alter the warm inflationary evolution.
\end{abstract}
\pacs{PACS number(s): }

\maketitle
\section{introduction}

Attempts to build inflationary models based on supersymmetric Grand Unified
Theories run into difficulties caused by the size of the supergravity
corrections to the inflaton potential. Inflation requires severe flatness
conditions on the potential, but these conflict with the $F$-term supergravity
corrections. The solutions to this problem have meant considering models with
special cancellations or models where a different supergravity correction, 
the $D$-term, dominates   (see \cite{lyth98} for a review).

A totally different solution to the problem of supergravity corrections has
recently been put forward, which is based on the realisation that the
dissipation associated with warm inflation relaxes the constraints on the
flatness of the potential \cite{berera03}. Warm inflation can exist in a
parameter regime where the supergravity corrections to the potential can be
safely ignored. 

In warm inflation, particle production during inflation provides a damping
effect on the inflaton. This idea has been around for a long time 
\cite{moss85,lonsdale87,yokoyama88,liddle89}, but the general features of this
scenario were described in \cite{berera95}. The non-equlibrium dynamics has
been expensively developed in subsequent work 
\cite{berera96,berera97,berera98,berera99,yokoyama99,moss02,
berera01,berera03,lawrie01} and several phenomenological warm inflation models
have been discussed in the literature 
\cite{oliveira98,bellini98,maia99,chimento02,oliveira02-2}.

Models in which warm inflation appears to occur spontaneously have the inflaton
decaying by two stages, the first stage into a heavy particle and the second
into a light particle \cite{berera03,Berera:2004kc}. An example is provided by
the interaction Lagrangian
density,
\begin{eqnarray}
{\cal L}_I = - g^2 \phi^2 \chi^2 - \frac1{\surd2}\,h \chi {\bar \psi_y}\psi_y
\label{lint}
\end{eqnarray}
where $\chi$ is a heavy boson and $\tilde y$ (field $\psi_y$) is a light
fermion. The dissipation is associated with $\phi\to\chi\to\tilde y\tilde y$. 

The ultimate destination of the vacuum energy from the inflationary phase is
into excitations of the light sector fields. In warm inflation, one has to
consider the possibility that these excitations enhance the loop corrections
to the inflaton potential and violate the  flatness conditions which inflation
requires. Because of this concern, we have calculated the loop corrections to
the potential under the assumption that the light fields thermalise.

The light sector will typically have coupling terms representing
self-interactions, or interactions with other light fields, in addition to the
couplings given in eq (\ref{lint}). The relaxation time of the radiation
should therefore be independent of the damping mechanism which is affecting
the inflaton. Whether the radiation thermalises during inflation is therefore
rather arbitrary. We assume thermalisation, but some consequences of
non-thermalisation are mentioned in the conclusion.
 
We shall consider the simplest inflationary models which include global
supersymmetry. These models divide naturally into two classes. In the first
class, which we call pure, the vacuum energy is associated only with the
inflaton field. We find that normalising the density perturbation amplitude to
the cosmic microwave background implies a mass scale of up to $10^{11}GeV$ and
coupling constants $g$ and $h$ around $0.1$.

In the second class of models, part of the vacuum energy can be linked to a
false vacuum of the $\chi$ field. These are the supersymmetric hybrid models
of inflation \cite{copeland94,dvali94}. We find that normalising the density
perturbation amplitude to the cosmic microwave background implies a
mass scale of up to $10^{14}GeV$ for the false vacuum energy and coupling
constants again around $0.1$ . Consequently, $F$-term supersymmetric inflation
with parameters in this range is of the warm inflationary type.

Whilst the pressent work was nearing completion, we learned that an independent
study of hybrid models of warm inflation was also underway, being conducted by
Arjun Berera and Mar Bastero-Gil \cite{Bastero-Gil:2004tg}.

\section{Supersymmetric Models}

\subsection{Potential and interaction terms}

We have just described how the warm inflationary scenario arises when there is
a two stage reheating process involving a heavy boson.  Global SUSY models can
easily be constructed which provide the required interactions 
\cite{berera03,Berera:2004kc}.
Consider the superpotential
\begin{equation}
W = -g \Phi X^2,
\end{equation}
where the scalar field components of the chiral superfields $\Phi$ and $X$ are 
$\varphi$ and $\chi$ respectively. 
The scalar  interaction terms in the theory are unpacked  from
the superpotential using
\begin{equation}
{\cal L_S} = -|\partial_\Phi W|^2-|\partial_XW|^2
\end{equation}
We identify the inflaton with $\phi=\sqrt{2}{\rm Re}\,\varphi$, and then
\begin{equation}
{\cal L_S} = -g^2|\chi|^4 -2g^2 \phi^2|\chi|^2
\label{lagp}
\end{equation}
Supersymmetry breaking now plays an important role in determining the shape of
the inflaton  potential along the flat direction $\chi=0$ \cite{lyth98}. A `new
inflation' type of model \cite{linde82,albrecht82,hawking82} results from
introducing a soft SUSY breaking mass $M_s$ for the $\chi$ field. The inflaton
potential $V(\phi)$ is determined by the one loop correction
\cite{Hall:2004ab},
\begin{equation}
V(\phi)={1\over 2}g^2M_s^2\left(\phi^2\log{\phi^2\over\phi_0^2}
+\phi_0^2-\phi^2\right)\label{ppot}
\end{equation}
Supercooled inflation requires $\phi_0>m_p$, but in this parameter range
the inclusion of supergravity $F$-term corrections would typically prevent the
inflation from occurring.

Hybrid inflationary models can be constructed  if we change the superpotential
slightly \cite{copeland94,dvali94},
\begin{equation}
W = g \Phi \Lambda^2-g\Phi X^2+g\Phi X^{\prime 2}.\label{hspot}
\end{equation}
where $\Lambda$ is a constant and $X$ and $X'$ are a pair of superfields.
The interaction terms are now
\begin{equation}
{\cal L_S} = -g^2|\chi^2-\chi^{\prime 2}-\Lambda^2|^2
 -2g^2 \phi^2(|\chi|^2+|\chi'|^2)
\label{lagh}
\end{equation}
In hybrid models, the $\chi$ field is stable at $\chi=0$ during inflation and
the potential is dominated by the constant term $g^2\Lambda^4$. The $\chi$
field becomes unstable at when $\phi$ falls below the critical value
$\phi_c=\Lambda$. 

The supersymmetry is broken by the inlaton field resulting in a non-vanishing
one loop contribution to the inflaton potential. The presence of the second
superfield helps produce a potential which is suitable for inflation by
reducing the size of the quantum corrections. For $\phi\gg\Lambda$, 
\begin{equation}
V(\phi)=g^2\Lambda^4+
{g^4\over4\pi^2}\Lambda^4\ln\left(2g^2\phi^2\over
\Lambda^2\right)\label{hpot}
\end{equation}
The heavy sector plays a double role in contributing to the vacuum energy and
damping the inflaton field.\footnote{For comparison with Dvali et al
\cite{dvali94}, their $\kappa=2g$ and $\mu=g^{1/2}\Lambda$.}

For an efficient two stage reheating process, we introduce an additional light
sector  $Y$, which can be coupled through a superpotential
\begin{equation}
W = -g \Phi X^2-h X Y^2\label{superhpot}
\end{equation}
The Yukawa interaction terms are recovered from
\begin{equation}
{\cal L_{Y}} = -\frac{1}{2}
\frac{\partial^2 W}{\partial \phi_n \partial \phi_m}
\overline\psi_n P_L \psi_m -
\frac{1}{2}
\frac{\partial^2 W^*}{\partial \phi_n^* \partial \phi_m^*}
\overline\psi_nP_R\psi_m 
\end{equation}
where $\phi_m$ is a superfield and $P_L=1-P_R=(1+\gamma_5)/2$. 
The interactions contain terms such as those in eqn. (\ref{lint}), and lead
to a friction term $\propto \dot\phi$ in the inflaton field equation
\cite{berera03}. They also have an effect on the vacuum polarization of the
$\chi$ field, which in turn can affect the inflaton potential. The full set of
interaction terms and the vacuum polarisation are discussed in  section 3. 

\subsection{Inflationary dynamics}

In an expanding, homogeneous universe, the inflaton equation of motion is given
by
\begin{equation}
\ddot\phi+(3H+\Gamma)\dot\phi+V_{T,\phi}=0,\label{wip}
\end{equation}
where  $V_T(\phi,T)$ is the thermodynamic potential and $\Gamma(\phi,T)$ is the
damping term due to interactions between the inflaton $\phi$ and
surrounding fields.  
For supercooled inflation, this damping term is negligible compared to the
Hubble damping term.  
The interesting regime of warm inflation is characterised by large damping
terms.  To distinguish between
the two inflationary scenarios, a dimensionless parameter, $r$, is introduced
to denote the relative strength between the damping terms
\begin{eqnarray*}
r={\Gamma\over 3H}.\label{rdef}
\end{eqnarray*}
We shall take warm inflation in the limit $r \gg 1$.

As the inflaton evolves, energy dissipates into radiation and entropy is
produced. Simple thermodynamic relations lead to a definition of entropy
density,
\begin{equation}
s(\phi,T)=-V_{T,T}={4\pi^2\over 90}g_*T^3+\dots,
\label{s}
\end{equation}
where $g_*$ is the effective particle number and the dots denote contributions
from the thermal correction to the potential. In the warm inflationary
scenario, inflation is characterised by three
slow-roll equations \cite{Hall:2003zp} 
\begin{equation}
\dot\phi=-{V_{T,\phi}\over 3H(1+r )},\qquad
Ts=r \dot\phi^2,\qquad
3H^2=8\pi GV_T.\label{slow}
\end{equation}
The second equation denotes conservation of energy, while the third is the
usual Friedmann equation.
Slow-roll automatically implies inflation, $\ddot a>0$, and the consistency of
slow-roll is governed by a set of slow-roll parameters:
\begin{equation}
\epsilon={m_p^2\over 16\pi}\left({V_{T,\phi}\over V_T}\right)^2,\quad
\eta={m_p^2\over 8\pi}\left({V_{T,\phi\phi}\over V_T}\right),\quad
\beta={m_p^2\over 8\pi}\left({\Gamma_{,\phi}V_{T,\phi}\over \Gamma
V_T}\right),\quad
\delta={TV_{T,\phi T}\over V_{T,\phi}},
\label{slowrp}
\end{equation}
where $m_p^{-2}$ is Newton's constant. The slow-roll approximation is
consistent when the above parameters are less
than $r$. Supergravity $F$-term corrections, without special cancelations, lead
to $\eta$ of order unity \cite{copeland94}. We shall therefore concentrate on
the range,
\begin{equation}
1<\eta<r.\label{sloweta}
\end{equation}
The thermal corrections to the potential will be calculated in the next
section. For the remainder of this section we shall examine the situation
where the corrections are small, $V_T=V$ and the slow-roll parameter
$\delta=0$.

An important observational constraint on the model is set by the density
perturbation amplitude. In our case, where we have assumed that the radiation
has thermalised, the thermal fluctuations induce scalar density fluctuations.
The amplitude $\Delta$ can be obtained analytically
\cite{taylor00,Hall:2003zp}, and for  $r\gg1$,
\begin{equation}
V_h^{1/4}=\alpha r_h^{-3/4}\epsilon_h^{1/4}\Delta^{2/3}\,m_p
\label{potpert}
\end{equation}
where $\alpha\approx0.68g_*^{-1/12}$ and the parameters are evaluated at the
time $t_h$ that the perturbation scale crossed the horizon. The value of
$\Delta$ infered
from cosmic microwave observations is around $\Delta\approx 5.4\times 10^{-5}$
on the $500$ MPc scale \cite{map03}.

Limits on the mass parameters can be found by combining the slow-roll
limits (\ref{sloweta})  with the constraint from the density perturbations
(\ref{potpert}).  The pure inflation model with potential (\ref{ppot}) has two
mass parameters $gM_s$ and $\phi_0$. Order of magnitude estimates can be
obtained by taking $\phi_h\sim\phi_0$ for the value of $\phi$ at horizon
crossing (which is  consistent with numerical solutions \cite{Hall:2004ab})
and $\epsilon\approx\eta$. The normalisation condition (\ref{potpert}) gives
\begin{equation}
gM_s\approx 3.3\times 10^{-6}\eta_h r_h^{-3/2}m_p.
\end{equation}
The upper limit for $gM_s$ set by (\ref{sloweta}) is of the order
$10^{13}$ GeV.

For hybrid inflation, a similar approximation can be made when the vacuum
energy $g^2\Lambda^4$ dominates the potential and $\phi\gg\Lambda$. In this
case,
\begin{equation}
\epsilon\approx{g^2\over 4\pi^2}\eta
\approx{g^4m_p^2\over 64\pi^4\phi^2}\label{epseta}
\end{equation}
The normalisation condition (\ref{potpert}) gives 
\begin{equation}
\Lambda\approx3.88\times10^{-4}\eta_h^{1/4}r_h^{-3/4}\,m_p\label{lamr}
\end{equation}
When combined with the conditions $r_h>\eta_h>1$, the upper limit on $\Lambda$
is of order $10^{15}$ GeV.

More detailed limits can be placed on the parameters when we know the form of
the friction term in the inflaton equation. For the interactions in eqn.
(\ref{lint}), the friction term has been calculated in the zero-temperature
limit \cite{Berera:2004kc} and is given
to leading order in $h$ by
\begin{equation}
\Gamma \approx  \gamma\phi.
\label{gamma}
\end{equation}
The value of $\gamma$ depends on the decay process. For $\chi\to2\tilde y$ and
$\chi\to 2y$, the leading order contributions to $\gamma$ are
\begin{equation}
\gamma(\chi\to2\tilde y)=\gamma(\chi\to 2y)=
\frac{\sqrt{2}\,g^3\,h^2}{128\pi^2}
\end{equation}
For the fermionic channel $\tilde\chi\to y\tilde y$,
\begin{equation}
\gamma(\tilde\chi\to y\tilde y)=
\frac{3\sqrt{2}\,g^3\,h^2}{64\pi^2}
\end{equation}
Hence the total
\begin{equation}
\gamma={\sqrt{2}g^3h^2\over 16\pi^2}\label{geq}
\end{equation}
To be in the perturbative regime, with $g<1$ and $h<1$, sets a requirement
$\gamma<8.9\times10^{-3}$. 

We can relate $r$ to $\gamma$ using the slow-roll
equations (\ref{slowrp}),
\begin{equation}
r={\Gamma\over 3H}={\gamma\phi \,m_p\over (24\pi V)^{1/2}}
\end{equation}
The non-hybrid models can be regarded as having three parameters $gM_s$,
$\phi_0$ and $\gamma$. The normalisation from the density fluctuation
amplitude provides one constraint which can be used to eliminate one
parameter, let's say $\phi_0$. The warm inflationary regime can then be
displayed as bounds on the remaining two parameters.

The horizon crossing timescale depends on the number $N$ of e-folds of the
scale factor before the end of inflation. As a rough guide, we can take
$r_h=N\eta_h$. The normalisation condition (\ref{potpert}) gives
\begin{equation}
\phi_0=8.2\times 10^{-6}\gamma^{-1}N^{-1/2}m_p\label{phin}
\end{equation}
Note that $\phi_0<m_p$ is needed for $\eta$ to remain larger than one
throughout the inflationary era, which is a necessary requirement for the
neglect of supergravity corrections. After eliminating $\phi_0$, there is a
consistency requirement
\begin{equation}
\gamma\,gM_s<4.8\times 10^{-11}N^{-2}m_p\label{gamm}
\end{equation}
for warm inflation.

The hybrid models can be taken to have parameters $M=g^{1/2}\Lambda$, $g$ and
$\gamma$. The horizon crossing timescale is given approximately by the relation
$r_h=3N\eta_h$. The normalisation condition (\ref{potpert}) allows us to
express $g$ as
\begin{equation}
g\approx 1.5\times 10^{21}\gamma^2M^2N^{5/2}m_p^{-2}
\end{equation}
The condition $\eta>1$ becomes
\begin{equation}
\gamma>1.3\times10^{-6}N^{-1/2}
\end{equation}
Taken together with the expression for $\gamma$ given by eq. (\ref{geq}), this
gives lower limits on $g$ and $h$. For example, if $g\approx h$, then $g>0.1$.
Another condition follows from $\phi>\Lambda$, which imples
a consistency condition
\begin{equation}
\gamma M>3.8\times 10^{-12}N^{-13/8}m_p
\end{equation}
for the warm inflationary regime.

\begin{center}
\begin{figure}[ht]
\subfigure[]{\scalebox{0.7}{\includegraphics{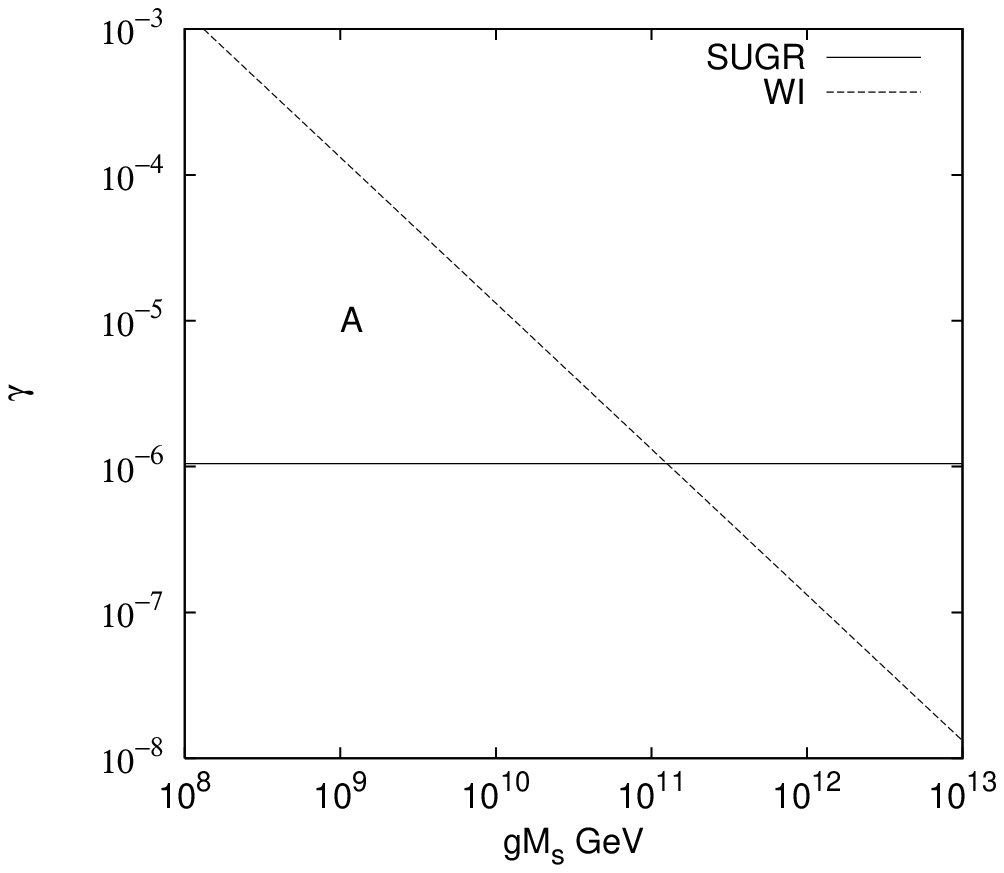}}}
\subfigure[]{\scalebox{0.7}{\includegraphics{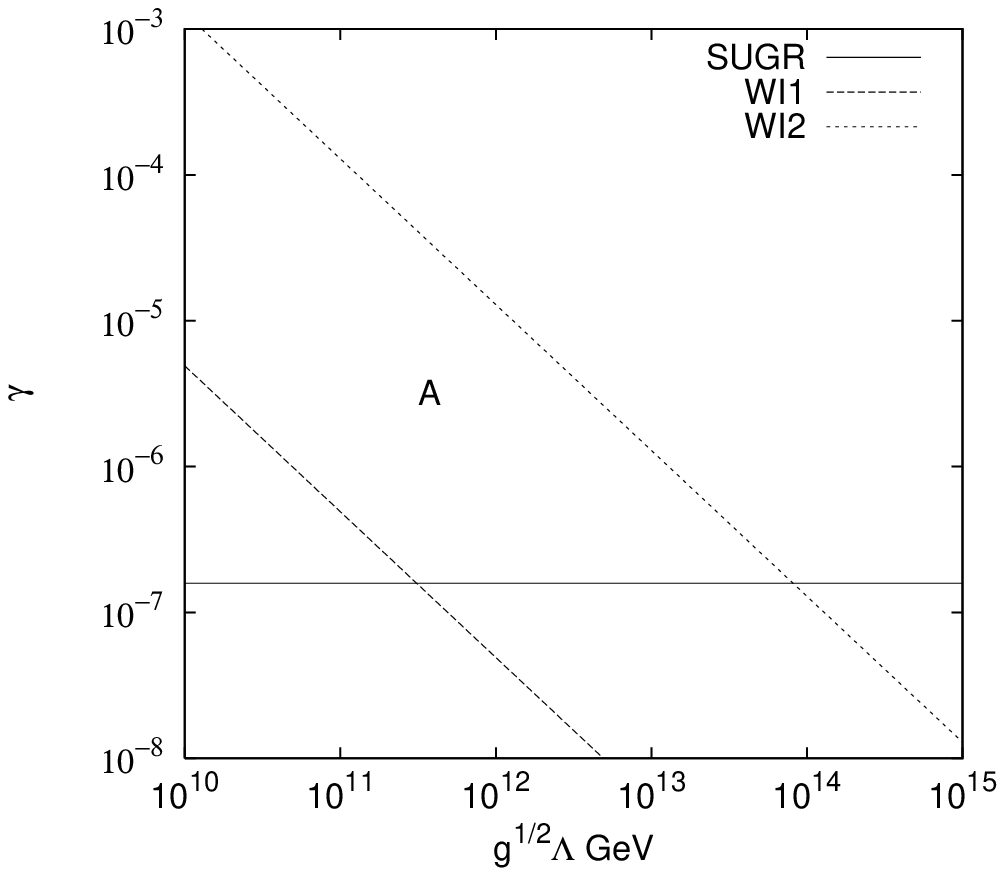}}}
\caption{The figures show, approximately, the allowed values (region $A$) of
the parameters  $\gamma$, $gM_s$ and $g^{1/2}\Lambda$ for (a) pure and (b)
hybrid inflation. The line SUGR shows the limit for which SUGR corrections can
be neglected. In case (a),  the limit of consistent warm inflation is shown as
WI. In case (b), the limits for consistent warm inflation are
$\phi_h=\Lambda$ (WI1) and $g=1$ (WI2).}
\label{fig1}
\end{figure}
\end{center}

The parameter ranges  are shown in figure \ref{fig1}. The number of e-folds of
inflation has been taken to be $N=60$. It
is clear from the figure, that the warm inflation occurs for a
broad range of parameters.  In the case of hybrid inflation, range includes
$0.03<g<1$ which means that many $F$-term inflationary models should be have
warm rather than cold inflation \cite{dvali94}.

Finally, we can use the slow-roll equations to determine what the temperature
is during the inflationary phase. The slow-roll equations (\ref{slow}) give
\begin{equation}
Ts=\frac23r^{-1}\epsilon V\label{tseq}
\end{equation}
Once again making use of the normalisation provided by the density fluctuation
amplitude, and the slow-roll relation
\begin{equation}
V_{,\phi\phi}={8\pi\eta V\over m_p^2},\label{vpp}
\end{equation}
we can compare $T_h$ to the slope of the potential,
\begin{equation}
T_h\approx230\,r_h^{1/2}\,\eta_h^{-1/2}\,(V_{,\phi\phi})^{1/2}
\label{temp}
\end{equation}
Note that $T_h$ is always larger than the mass scale responsible for the slope
of the potential. In this situtation, we should be concerned that thermal
corrections to the potential may make the models untenable. The calculation of
these thermal corrections to the potential is therefore necessary.

In the case of pure inflation, $V_{,\phi\phi}\sim 2g^2M_s^2$ and the
temperature during inflation is approximately
\begin{equation}
T_h\sim 320\,N^{1/2}\,gM_s.\label{tempp}
\end{equation}

In the case of hybrid inflation, using eq. (\ref{tseq}) and eq. (\ref{epseta}),
\begin{equation}
T_h\approx\left({45\over 16\pi^5}\right)^{1/4}g^{1/2}N^{-1/4}M\label{temph}
\end{equation}
Note that the vacuum energy of the hybrid model after inflation, before the
second field decays, is approximately $M^4$. The values of the temperature are
sufficiently high such that, if we have local supersymmetry, then we are in
danger of violating constraints set by the thermal producation of gravitinos 
\cite{Weinberg:1982zq,Ellis:1984er,Ellis:1990nb,Kawasaki:1994af,Kallosh:1999jj,Lyth:1999ph}. 
These constraints can be satisfied by taking small values of the mass
parameters, corresponding to large values of $r$ \cite{taylor01}. The allowed
parameter ranges permit this, but so far this appears to be an unnatural
feature of the models under discusion.

\section{Thermal Corrections}

The full set of interaction terms obtained from the superpotential
(\ref{superhpot}) are
\begin{eqnarray}
{\cal L_S}& = &g^2(\Lambda^2-|\chi|^2)^2 +4g^2 |\varphi|^2|\chi|^2
+4h^2|y|^2|\chi|^2+h^2|y|^4
+2gh(y^2\varphi^\dagger\chi^\dagger+y^{\dagger2}\varphi\chi)
\label{lagrangian2}\\
{\cal L_{Y}} &= &g(\varphi \overline \psi_{\chi} P_L\psi_{\chi}+
\varphi^\dagger \overline \psi_{\chi}P_R \psi_{\chi})
+h(\chi \overline\psi_{y}
P_L\psi_{y} +  \chi^\dagger \overline\psi_{y} P_R \psi_{y})\nonumber\\
&&+2g(\chi\overline\psi_{\chi}P_L \psi_{\varphi} +
\chi^\dagger\overline\psi_{\chi} P_R \psi_{\varphi}) 
+2h(y\overline\psi_{y} P_L\psi_{\chi} +  y^\dagger\overline\psi_{y}
P_R\psi_{\chi}) .
\label{lagrangian}
\end{eqnarray}
Thermalisation conditions of the light fermion, $\tilde y$ depend directly on
the mass of the fermion and its self-interaction.  These properties are
inherent to quadratic and cubic terms in the superpotential ($\mu_{\chi}Y^2$,
$\lambda Y^3$), which we have not specified.   We assume that the interactions
are such
that the light fermions thermalise and we calculate the corresponding thermal
effects.  The $\psi_y$ interactions to fields other than the inflaton, $\phi$
or $\chi$ fields, will have no effect on the thermal corrections to the
inflaton effective potential.  We can therefore disregard the exact nature of
these interactions.

For the model considered, if $\tilde y$ thermalises but $\chi$ and
$\psi_{\chi}$ do not, then important simplifications can be made.  The thermal
corrections in the action appear as a result of the self-energies of the
$\chi$ and $\tilde \chi$ fields.  Inside the self-energy loops, $y$ and
$\tilde y$ are taken to be very light, so the Hard Thermal Loop (HTL)
approximation can be made, i.e. $T \gg m_y, m_{\tilde y}$. Outside the loop,
however, the $\chi$ fields are heavy, and $T\ll m_{\chi}, m_{\tilde\chi}$.  

We use the imaginary time formalism and adopt the notation that 4-momenta are
written in upper case and 3-momenta as
written in lower case, so that $P^{\mu} = (\omega, {\bf p})$. The boson and
fermion propagators of the $\chi$ fields are $G$ and $S$ respectively. The
contribution to the effective potential of the inflaton field from the $\chi$
fields is given by
\begin{equation}
V_\chi=\int {d^4P\over(2 \pi)^4} \ln~\det \left(G^{-1} \right) - 
\int {d^4P\over(2 \pi)^4} ~\ln~\det \left( S^{-1}
S^{*-1}\right)^{1/2}\label{efpot}
\end{equation}
after regularisation has been applied. The detailed calculation of the thermal
corrections to the fermionic and  bosonic masses $m_\chi$ and
$m_{\tilde\chi}$ will be given in the following
sections.  

\subsection{Fermion Contribution}

\begin{center}
\begin{figure}[ht]
\scalebox{1.0}{\includegraphics{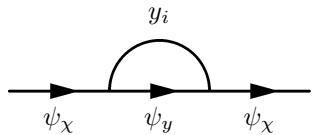}}
\caption{Diagram for the fermionic self energy}
\label{fermionloop}
\end{figure}
\end{center}
     
If we set $y=(y_1+iy_2)/\sqrt{2}$, then the $y_i \overline \psi_y \psi_{\chi}$
terms in eqn. (\ref{lagrangian}) lead to two similar fermion self-energy
diagrams  (fig. \ref{fermionloop}) for $\tilde\chi$. The vertex factors are
$i\sqrt{2}h$ and $\sqrt{2}h\gamma_5$ respectively.  Using the properties of
$\gamma_5$, contributions from both diagrams are found to be identical. 
Thermal feynman rules applied to the diagram in
figure \ref{fermionloop} lead to an expression for the fermion self-energy,
$\Sigma$,  
\begin{eqnarray}
\Sigma(P) &=& -4 h^2\,T\,\sum_n\int \frac{d^3k}{(2 \pi)^3}
 (K\!\!\!\!/ - P\!\!\!\!/) \Delta(K) \widetilde{\Delta}(P-K).
\end{eqnarray}
where $\Delta(K)\approx K^{-2}$, $k^0=2n\pi T$ for bosons and $k^0=(2n+1)\pi T$
for fermions (denoted by a tilde).

Many similar diagrams appear in the literature, for example
\cite{Klimov:1981ka,Weldon:1982bn,lebellac,Thoma:1995yw}, specifically with
regards to the HTL loop corrections to electron propagators in QCD.  The
calculation given here follows \cite{lebellac} with only slight
changes, due to having a scalar field rather than a vector field. Therefore we
only need quote the result,
\begin{equation}
\Sigma(P) = \frac{m_f^2}{2p} \gamma_0 Q_0 \left( \frac{i\omega}{p}\right) +      
\frac{m_f^2}{2p} \gamma \cdot \widehat{p}\left[1- \frac{i\omega}{p} Q_0
\left(      \frac{i\omega}{p}\right) \right]
\end{equation}
Note that the overall factor of $1/2$, which is different to the literature, is
a convention we have adopted for the definition of $m_f$ for later convenience.
$Q_0(x)$ is the Legendre function of the second kind,
\begin{equation}
Q_0(x) = \frac{1}{2} \ln \frac{x+1}{x-1}
\end{equation}
Accounting for the contribution of two diagrams, the fermion thermal mass is
\begin{eqnarray}
m_f^2 = \frac{h^2T^2}{2}.
\end{eqnarray}
In our conventions, the inverse propagator is 
$iS^{-1} = P\!\!\!\!/-m_{\tilde\chi}-\Sigma$
and thus can be written
\begin{eqnarray}
i S^{-1} = A_0\gamma_0 - A_s\gamma \cdot \widehat{p} -m_{\tilde\chi} 
\end{eqnarray}
where
\begin{eqnarray}
A_0 &=& i\omega - \frac{m_f^2}{2p} Q_0 \left( \frac{i\omega}{p}\right)\\
A_s &=& p + \frac{m_f^2}{2p} \left[1-\frac{i\omega}{p}Q_0
\left(                    \frac{i\omega}{p}\right)\right]
\end{eqnarray}
Hence the combination
\begin{equation}
(SS^*)^{-1}=-A_0^2+A_s^2+m_{\tilde\chi}^2
\end{equation}
The fermionic contribution to the effective potential (\ref{efpot}) becomes
\begin{equation}
V_{f} =   -2 \int \frac{d^4P}{(2 \pi)^4} \ln 
\left[ \left( \omega + \frac{i  m_f^2}{2p}Q_0\left( 
\frac{i\omega}{p} 
\right) \right)^2 +
 \left( p +  \frac{m_f^2}{2p} -\frac{i m_f^2}{2p^2}\omega Q_0
\left( \frac{i\omega}{p} \right)\right)^2 + m_{\tilde\chi}^2\right].
\label{vfexact}
\end{equation}
We can obtain the leading terms using $m_\chi\gg m_f$,
\begin{equation}
  V_{f}  = -2 \int \frac{d^4P}{(2 \pi)^4} \ln \left\{ \omega^2 + p^2
+m_{\tilde\chi}^2 +     
m_f^2\right\} + O(m_f^4)
\end{equation}
The evaluation of this regularised integral is a standard excercise, 
\begin{equation}
V_{f}    = -\frac{1}{32\pi^2} \left(m_{\tilde\chi}^4 + 2m_{\tilde\chi} m_f^2
\right) 
\ln \left(\frac{m_{\tilde\chi}^2 +             
m_f^2}{\mu^2} \right)
\end{equation}
where $\mu$ has been introduced by the regularisation. 

\subsection{Boson Contribution}

\begin{center}                                                                             
\begin{figure}[ht]
   \begin{center}
   \mbox{ 
   \subfigure[]{
\scalebox{1.0}{\includegraphics{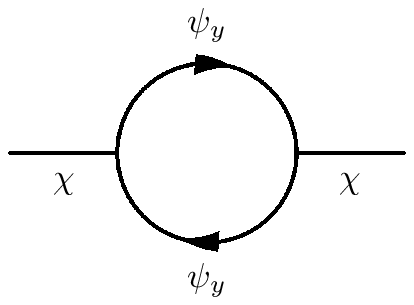}}\label{bosona}
} \quad
   \subfigure[]{
\scalebox{1.0}{\includegraphics{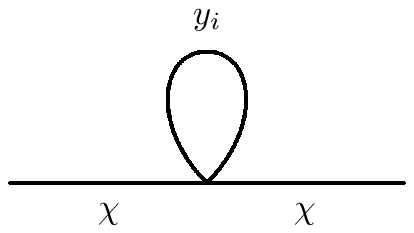}}\vspace{10cm}\label{bosonb}
}
\quad
   \subfigure[]{
\scalebox{1.0}{\includegraphics{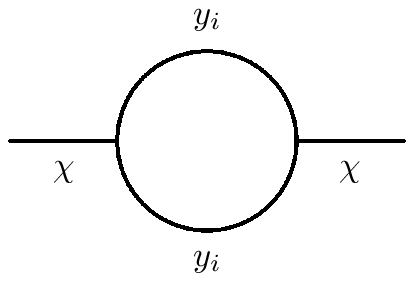}}\label{bosonc}
}
   }
   \caption{Diagrams for the bosonic self energy}
   \label{bosonloop}
   \end{center}
\end{figure}
\end{center}

Three diagrams are expected to contribute to the bosonic self-energy as shown
in figure \ref{bosonloop}.  Due to the interaction terms $\chi_{i}
\overline\psi_{y} \psi_{y}$ in eqn. (\ref{lagrangian}), there will be two
diagrams similar to figure \ref{bosona}, with vertex factors $i\sqrt{2}h$ and 
$\sqrt{2}h\gamma_5$ respectively. The two diagrams result in identical
expressions, each with a symmetry factor of $1/2$ because the fermions
are Majorana. The self-energy for $\chi_1$ from figure \ref{bosona} is given
by 
\begin{equation}
\Pi(P)_a = h^2\,T\,\sum_n\int \frac{d^3k}{(2 \pi)^3}
Tr\left[K\!\!\!\!/(K\!\!\!\!/ - P\!\!\! \!/)\right] 
\widetilde{\Delta}(K) \widetilde{\Delta}(K-P) 
\end{equation}
In the HTL limit,
\begin{equation} 
  \Pi(P)_a= -4 h^2  \,T\,\sum_n\int \frac{d^3k}{(2 \pi)^3}K^2
\widetilde{\Delta}(K) \widetilde{\Delta}(K-P) 
\end{equation}
Using $\widetilde{\Delta}(K) \approx K^{-2}$ the final result is a single
fermion loop expression,
\begin{equation}
\Pi(P)_a =-4h^2\,T\,\sum_n\int \frac{d^3k}{(2 \pi)^3}
\widetilde{\Delta}(K) =\frac16h^2T^2.
\end{equation}
The second contribution stems from the four tadpole diagrams coming from the
terms $\chi_i^2 y_i^2$ in eqn. (\ref{lagrangian2}).
Each diagram is identical, and has a symmetry factor of $1/2$. The self energy
is given by the expression,
\begin{equation}     
\Pi(P)_b = 4 h^2\,T\,\sum_n\int \frac{d^3k}{(2 \pi)^3}
{\Delta}(K) 
  =  \frac13h^2T^2
\end{equation}

The Feynman diagram in figure \ref{bosonc} does not result in an $O(T^2)$
contribution to the self-energy, but instead gives a possibly important
contribution of  $O(m_\chi^2)$. The
self-energy contribution is given by
\begin{eqnarray}
\Pi(P)_c &=& 4g^2 h^2\phi^2\,T\,\sum_n\int \frac{d^3k}{(2 \pi)^3}
 {\Delta}(K) {\Delta}(K-P)
\end{eqnarray}
For the evaluation of this integral in the HTL limit see, for example,
\cite{Andersen:1997hq}. The result is
\begin{equation}
\Pi(P)_c = 4g^2 h^2\phi^2\int{dk\over (2\pi)^2}
{1\over p}\ln\left({k-p\over k+p}\right){1\over e^{kT}-1}.
\end{equation}
The integral evaluuates to
\begin{equation}
\Pi(P)_c \approx {1\over 2\pi^2}g^2 h^2\phi^2\log{T^2\over p^2}
\end{equation}
in the HTL limit.

The inverse boson propagator is defined as $G^{-1} = P^2 + m_\chi^2 +\Pi$. 
The total contribution from the first two diagrams in figure \ref{bosonloop}
defines a contribution $m_b^2$ to the mass,
\begin{eqnarray}
m_b^2\equiv \Pi_a+\Pi_b= \frac12h^2T^2 
\end{eqnarray}
Note that $m^2_\chi=4g^2\phi^2$,  and the temperature dependent part of $\Pi_c$
can be regarded as a contribution to the finite temperature coupling constant
$g(T)$,
\begin{equation}
g^2(T)=g^2\left(1+{1\over 8\pi^2} h^2\log{T^2\over\mu^2}\right).
\end{equation}
Since the inflationary dynamics is not sensitive to the precise value of $g$,
we shall not distinguish between $g$ and $g(T)$.

The bosonic contribution to the effective potential is obtained from the two
fields $\chi_i$,
\begin{eqnarray}
V_{b} &=& \int \frac{d^4P}{(2 \pi)^4} \ln \left\{ P^2 +m_\chi^2 + m_b^2
\right\}         \nonumber  \\
       &=& \frac{1}{32\pi^2} \left(m_\chi^2 + m_b^2 \right)^2 \ln
\left(\frac{m_\chi^2 +
m_b^2}{\mu^2} \right)
\end{eqnarray}

\subsection{Effective Potential} 

The total effective potential is given by the sum of both the fermionic and
bosonic contributions. The largest of the $O(T^2)$ terms cancel due to the fact
that $m_b=m_f$.  This is, of course, due to the underlying supersymmetry, but
it
happens despite the fact that supersymmetry is broken at non-zero
temperatures. The remaining $O(T^2)$ terms are due to other sources of SUSY
breaking. For the case of soft SUSY breaking, we can consider the $\chi$ boson
and
$\tilde\chi$ fermion masses,
\begin{eqnarray}
m^2_\chi&=&2g^2\phi^2+M_s^{\prime 2},\\
m^2_{\tilde\chi}&=&2g^2\phi^2.
\end{eqnarray}
The leading order terms in the potential are
\begin{equation}
V_\chi =\frac{1}{32\pi^2} \left\{
       \left(m^2_\chi + m_b^2 \right)^2 \ln \left(
\frac{ m^2_\chi + m_b^2}{\mu^2}     \right) 
      -\left(m^2_{\tilde\chi} + m_f^2 \right)^2 \ln
\left(\frac{m^2_{\tilde\chi} + m_f^2}{\mu^2} 
\right) \right\} + constant .
\end{equation}
For our scenario, $g\phi\gg M'_s$, which results in 
\begin{equation}
V_\chi  \approx
\frac12M_s^2\left(g^2\phi^2+\frac14h^2 T^2\right)
\left(\ln\left({g^2\phi^2+\frac14h^2 T^2\over
g^2\phi_0^2}\right)-1\right)
+\frac12g^2M_s^2\phi_0^2.\label{ptt}
 \label{effpot}
\end{equation}
We have defined $M_s^2=M_s^{\prime2}/(8\pi^2)$, and $\phi_0$ has been chosen
such that both the potential and its derivative vanish at $\phi = \phi_0$ when
$T=0$.

\begin{center}
\begin{figure}[ht]
\includegraphics{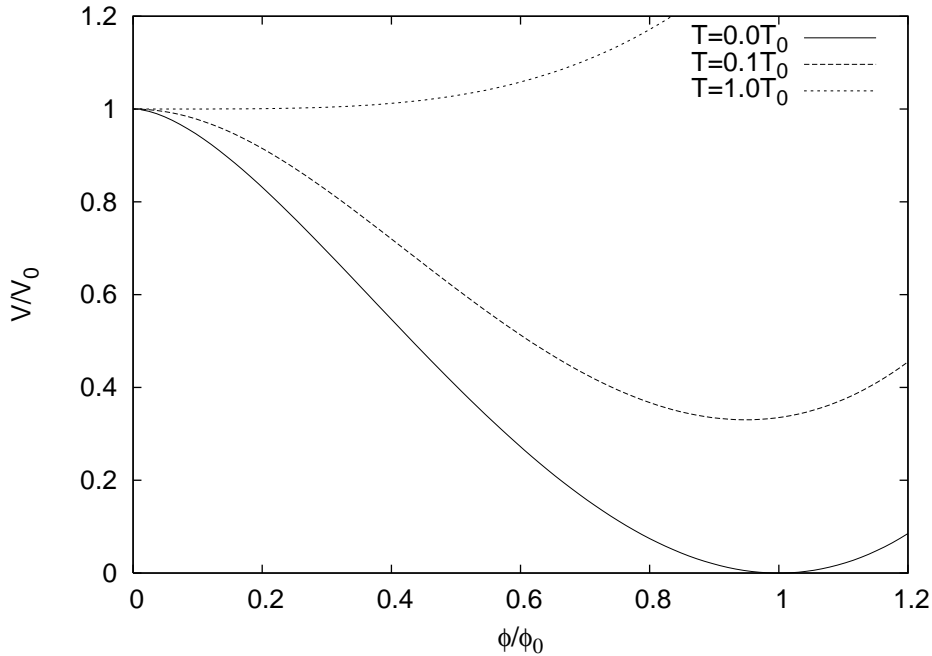}
\caption{The thermodynamic potential of the pure inflationary model is
depicted. On the vertical axis, $V=V_T(\phi,T)-V_T(0,T)$ and $V_0=V_T(0,0)$.
The critical temperature $hT_0=2g\phi_0$.}
\label{fig2}
\end{figure}
\end{center}

The thermodynamic potential of the inflaton in the models under consideration
is determined  primarily by the $\chi$ loop contribution calculated above and
the free energy of the light radiation fields,
 \begin{equation}                                                              
V(\phi,T) = -\frac{\pi^2}{90} g_{*}T^4 + V_\chi(\phi, T)               
\end{equation}
Figure \ref{fig2} graphically shows the temperature dependance of the        
potential.  The scale of the temperature corrections is set by $T_0$, where
$hT_0=2g\phi_0$.

The Hybrid models (\ref{hspot}) can be treated in a similar manner if we couple
the two heavy superfields to the light superfield with equal couplings $h$
(which is consistent with an underlying $U(1)$ symmetry).
The masses of the $\chi$ and $\chi'$ fields are
\begin{eqnarray}
m^2_{\chi}&=&2g^2\phi^2+2g^2\Lambda^2\\
m^2_{\chi'}&=&2g^2\phi^2-2g^2\Lambda^2\\
m^2_{\tilde\chi}&=&m^2_{\tilde\chi'}=2g^2\phi^2
\end{eqnarray}
The one loop correction to the inflaton potential is now
\begin{equation}
V_\chi={1\over 4\pi^2}g^4\Lambda^4\ln\left(
{2g^2\phi^2+\frac12h^2T^2\over g^2\Lambda^2}\right)\label{htpot}
\end{equation}
for $\phi\gg\Lambda$.

\subsection{Friction term}

The zero temperature friction term calculated by Berera and Ramos
\cite{Berera:2004kc} can be modified for our interaction terms with the decays
$\phi\to\chi\to2\tilde y$ and $\phi\to\chi\to2 y$,
\begin{equation}
\Gamma = {
g^4 \phi^2 \Gamma_{\chi} 
\over 2\pi \left(m^2_{\chi}+\Gamma^2_{\chi}\right)^{1/2}
\left(2m_{\chi}
\left(m^2_{\chi}+\Gamma^2_{\chi}\right)^{1/2}+2m^2_{\chi}
\right)^{1/2}
}
\end{equation}
where $\Gamma_\chi$ is the $\chi$ decay width,
\begin{equation}
\Gamma_\chi={h^2\over 16\pi}m_\chi
\left(1-{4m_y^2\over m_\chi^2}\right)^{3/2}
+{h^2\over 16\pi}m_{\chi}
\left(1-{4m_{\tilde y}^2\over m_{\chi}^2}\right)^{1/2}
\end{equation}
This reduces to the expression used earlier when $m_y=0$ and
$m^2_\chi=2g^2\phi^2$.

One would expect that the thermal corrections to $\Gamma$ will manifest
themselves as corrections to $m_{\chi}$ and $\Gamma_{\chi}$.  
Since the latter already contains factors of $h$, the thermal corrections to
$\Gamma$ will be of order $h^3~T^2$.  
The correction to $\Gamma$ due to $m_{\chi}$ will be of order $h^4~T^2$. 
The corrections to the effective potential are order $h^2~T^2$ and are
therefore taken to be the dominant effect.  
Corrections to the friction term can therefore be ignored.

\subsection{Effects on the inflationary dynamics}

The temperature range relevant for the warm inflationary scenario was given in
eq (\ref{temp}). In the models under discussion, $h^2T^2\ll g^2\phi^2$ and the
thermal corrections make only a small change to the height of the
thermodynamic potential $V_T$. The effect on the slope of the potential is
more delicate, however, and has to be investigated seperately. The slope can
be quantified by the slow roll parameter $\eta$. Consider the change in $\eta$,
\begin{equation}
{\delta\eta\over\eta}={V_{T,\phi\phi}-V_{,\phi\phi}\over V_{,\phi\phi}}
\end{equation}
In the non-hybrid case,
\begin{equation}
{\delta\eta_h\over\eta_h}\sim{h^2T_h^2\over g^2\phi_0^2}
\end{equation}
Given eqs. (\ref{tempp}),  (\ref{phin}) and (\ref{gamm}), this correction is of
order  $10^{-5}h^2g^{-2}$ at most.

For hybrid inflation, using (\ref{htpot}),
\begin{equation}
{\delta\eta_h\over\eta_h}\approx-{h^2T_h^2\over g^2\phi_h^2}
\end{equation}
Proceeding as above, using eq. (\ref{temph}), this correction
is of order $10^{-7}h^2g^{-2}$.

\section{Conclusion}

We have found that the inflation occurs naturally in particle models with
global
supersymmetry when the dissipative effects of particle production are taken
into account. The warm inflationary scenario escapes the flatness problems
which arise
when supercooled inflation is combined with global supersymmetry. The
parameter restrictions on the model are not severe, with the possible exception
of a gravitino constraint, and there is a correspondence between mass
parameters required for the observed density fluctuation amplitude and the
parameter values of interest for supersymmetric Grand Unified Theories.

We have demonstrated that,  in a two stage reheating process, the thermal
corrections to the inflaton potential are small, due to fermion-boson
cancellations . The assumptions used for the models have been relatively mild,
consisting mainly of the following:
\begin{itemize}
\item At least one superfield has vanishing, or very weak coupling, to the
inflaton and another has non-vanishing coupling. During inflation, the former
will naturally become a `light' sector, and the latter a `heavy' sector.
\item There is either (a) soft SUSY breaking in the heavy sector, which we
called the pure inflation model or (b) a false vacuum energy, and two equally
coupled heavy superfields, which we identify as the hybrid model.
\item We have assumed that the light radiation thermalises.
\end{itemize}
The last assumption was needed to avoid far from equilibrium calculations.
However, in the absence of thermalisation, the energy will still be dumped
into the light
sector. The important features of the light particle distribution can be
described in terms of non-thermal occupation numbers $n(k)$. If the boson and
fermion occupation numbers are similar, we might still expect the cancellation
of correction to the effective potential which we have found here. 

We have made use of the density fluctuation amplitude when setting limits on
the parameters in the models. If the radiation does not thermalise, we would
expect to find changes in the predicted value of the density fluctuation
amplitude and this remains to be investigated further.

\acknowledgements
We are grateful to Arjun Berera for discussions about this work and to Mar
Bastero-Gil for pointing out ref. \cite{dvali94}.

\bibliography{paper.bib}

\end{document}